# A non-scanning atomic magnetometer based on coherent population trapping

S. Pradhan[*], S. Mishra, and A.K. Das

Laser and Plasma Technology Division

Bhabha Atomic Research Centre, Mumbai-85, India

**Abstract**

The method of operation of an atomic magnetometer based on coherent population trapping (CPT) without any requirement of radio frequency scanning is demonstrated. Using a hybrid approach comprising of polarization rotation and tailored transmission due to CPT states, the simultaneous operation of the device as atomic frequency standard as well as atomic magnetometer is investigated. The magnetometer sensitivity can be further improved by utilizing the inherent option of magnetic field assisted synchronized radio frequency scanning.

[*]Corresponding author email: spradhan@barc.gov.in, spradhan75@rediffmail.com

Atomic devices based on laser-atom interaction have promising prospect in technological applications [1]. Particularly, the coherent population trappping (CPT) based atomic clock and atomic magnetometer has unique advantages due to possible miniaturization to chip scale [2,3]. This has been feasible due to elimination of microwave cavity and operation at room temperature in contrast to stringent requirement by the conventional atomic frequency standard and state of art SQUID magnetometer operating at cryogenic temperature respectively. However, the CPT based magnetometer has few severe drawbacks which limits its prospect to replace other atomic based or solid state magnetometers [4]. One of the principal limitation arises due to measurement of frequency difference between the field sensitive and field insensitive transition.

The CPT based atomic clock is realized by stabilizing the radio frequency (RF) oscillator to the magnetic field insensitive transtion. Generally, a small magnetic field is applied to separate the field insensitive transition form the field sensitive transition. The limitation of the transit time broadening is circumvented by filling the atomic cell with suitable buffer gas at desired vapor pressure. This method has been sussessfully utilized for realizing chip scale CPT based atomic clock. The other promising method of improving the coherent decay time is by using anti-relaxation coated cell or hybrid technique involing both buffer gas filled cell with an anti-relaxation coated cell enclousure [5,6]. In distinction to the CPT based atomic frequency standard, the CPT based atomic magnetometer requires measurement of the separation between the field sensitive and field insensitive transition. In presence of a bias magnetic field, any change in the magnetic field is reflected in the corresponding change in position of the field insensitive transition. It is not only cumbersome in determining the field direction, but also there is an ambiguity in measurement of field amplitude.

Recently, the intriguing aspects of the polarization rotation and transmission signal due to CPT states has been studied in the high field regime characterized by the characteristic width of the CPT resonances smaller than the Larmor's frequency [7]. Also, it has been demonstrated that in the low field regime (CPT width larger than Larmor's frequency), magnetic field amplitude as well as orientation along the laser propagation direction can be measured unambiguously [8]. This method rely on the shift between the magnetic field insensitive transmission due to CPT states and field sensitive polarization rotation signal. Based on this observation, the possible operation of a dual purpose atomic device for atomic frequency standard and magnetic field measurement is described [9]. In the dual purpose atomic device, the RF is stabilized to the centre of transmitted signal by the CPT states and the magnetic field is measured from the polarization rotation signal. The experiments were carried out with atomic cell without any buffer gas. The sensitivity of the atomic magnetometer and the precision of atomic frequency standard can be improved by using atomic cell filled with buffer gas, cell with anti-relaxation coating or a hybrid cell [6]. Though the later approach seems most promising, our investigation is limited to atomic cell in buffer gas envoirment as it has been sussessfully used in the chip scale atomic devices [2,3]. For comparision, we also present the data obtained with buffer gas free atomic cell.

The experimental procedure is quite analogous to typical VCSEL based CPT set-up and a schematic diagram is illustrated in figure-1 [10-12]. The majority part of the set-up is described in our earlier work [7-9], except few changes incorporated here. The VCSEL diode laser at 795 nm is modulated at ~1.708 GHz and the ±2 side bands are used for realizing quantum interference in $^{87}$Rb atoms in presence of $N_2$ gas at ~25 Torr. The desired imbalance in the orthogonal circular polarization component is introduced by passing the frequency modulated beam through a polarizing beam splitter cube (PBS) followed by a quarter wave plate at ~$8^0$. This is incontrast to ~$1^0$ used in our earlier experiment involving atomic cell without any buffer gas. The imbalanced bichromatic light field is passed through the $^{87}$Rb atomic cell placed in a magnetic field controlled environment. The magnetic field along and transverse to the laser propagation direction are controlled by three pairs of coils in orthogonal directions enclosed in several layers of magnetic field shield. After interaction with the atomic sample, the transmission and polarization rotation signal is monitored with the help of PBS, photo diodes and signal processing units involving lock-in amplifiers. The laser light is locked to the desired single photon detuning by imposing a low frequency modulation to the laser current, phase sensitively detecting the transmitted signal by PBS and using a servo loop. The CPT signals are phase sensitively detected with respect to a low frequency modulation (at a distinct frequency to that used for single photon spectroscopy) applied to the voltage controlled oscillator (VCO) frequency. The transmission and polarization rotation signal by the CPT states are extracted from the transmitted and the reflected light by the PBS respectively. The VCO frequency is locked to a synthesized signal generated from the transmission and polarization rotation signal. The change in the polarization rotation signal under the VCO locked to the synthesized signal is monitored with respect to the applied magnetic field.

In a buffer filled atomic cell, the pressure broadening of the excited state severely spoils the contrast of the CPT resonances obtained with linearly polarized light [6]. This can be partially circumvented by using $^{87}$Rb D1 transition (@ 795 nm) due to large separation of the excited states. It is interesting to note that the polarization rotation signal is relatively immune to such degradation. The polarization rotation signal has inherently high SNR as it is extracted from almost zero background in comparison to the transmission due to CPT states. As the polarization rotation signal is insensitive to magnetic field for perfect linearly polarized light, it can play an important role for atomic frequency standard. We would also like to point out that, near zero field the convoluted signal due to CPT states are enhanced for both atomic cell filled without buffer gas [8] or with buffer gas in sharp contradiction to the observation in ref-6. The degradation of CPT contrast near zero field in the said reference may be due to use of buffer gas filled cell with anti-relaxation coated cell enclosure. However, we observe that presence of small transverse field or workings in the medium field regime where resonance width are comparable to the Larmor's frequency degrade the CPT contrast significantly.

The amplitude of the polarization rotation signal gets significantly enhanced by increasing the degree of imbalance between orthogonal circular polarization components. This observation is consistent with the earlier experiment carried out in atomic cell without buffer gas [8]. Here we have used a relatively larger angle of the waveplate at ~$8^0$ in comparison to our earlier work. This is because larger amplitude of the polarization rotation signal is favorable in the present technique in contrast to the zero crossing of the polarization rotation playing important role in the prior art. The larger amplitude of the polarization rotation signal at higher imbalance is obtained with a compromise of magnetic field dependent distortion in the transmission signal profile as illustrated in figure-2. This signal distortion is also more prominent due to narrower line width of the transmission signal obtained in the buffer gas environment. From the insert in the fig-2, it is clear that the transmission profile due to CPT states does not have any unique position that is independent of magnetic field. Therefore, this signal is not suitable for atomic frequency standard. The distortion in the transmission signal profile is attributed to increased optical pumping as the imbalanced is increased and the inevitable reflection of significant CPT signal by the PBS [7,13,14]. Both of these effects are expected to be enhanced at higher imbalance between the orthogonal circular polarization components.

It may be noted that the polarization rotation signal due to CPT states has strong dependence on the magnetic field and is consistent with the prior art involving experiment in buffer gas free environment. The signal distortion of the transmission signal can be compensated by synthesizing a signal by combining the transmission and polarization rotation signal due to CPT states. The synthesized signal is shown in figure-3 along with polarization rotation signal due to CPT states. The synthesized signal for various magnetic field merge at a single two photon detuning position, which can be used for atomic frequency standard. Therefore, with VCO stabilized to the magnetic field insensitive position, the polarization rotation single can be monitored for measurement of magnetic field.

The VCO frequency locked to the magnetic field insensitive position of the synthesized signal along with polarization rotation signal are illustrated in figure-4 for various magnetic fields. It may be noted that the polarization rotation signal changes its amplitude and polarity depending on the amplitude and orientation of the magnetic field along the laser propagation direction. Therefore, the magnetometer is sensitive to both amplitude and orientation of the magnetic field. As demonstrated here, the measurement of magnetic field does not require any scanning operation across the two photon resonance as has been desired in conventional CPT based magnetometer. The noise in the polarization rotation signal is primarily due to the fluctuation in the locked signal. Since frequency stabilization is carried out on the synthesized signal, the relatively noisy transmission signal due to CPT states is the limiting factor for the sensitivity of the magnetometer.

The variation of the polarization rotation signal due to CPT states under the VCO locked to the magnetic field insensitive position of the synthesized signal (or transmission signal) is illustrated in figure-5. The experimental data shown in figure-5 (a) is a comparison between the $^{85}$Rb atoms without any buffer gas and $^{87}$Rb atoms with $N_2$ buffer gas at 25 Torr in atomic cell. The parameters for the $^{85}$Rb atoms are same as given in ref-8. Briefly, the D-2 transition at 780nm is used and the laser current is modulated at the half of the ground state hyperfine separation (1.517 GHz). The ±1 side bands are used for quantum interference. The VCO is stabilized to the CPT signal (in contrast to the combined signal for buffer gas filled cell). The amplitude of the polarization rotation signal is found to be varying linearly up to ~±1.8 µT, which constitutes the dynamic range of the magnetometer. Nevertheless, the signal amplitude is found to be varying proportionally up to ~±3.3 µT. The amplitude of noise in the polarization rotation signal correspond to ~25 nT and can be improved by using narrower resonance width, higher SNR and improving lock accuracy. The data for $^{87}$Rb atoms with $N_2$ buffer gas at 25 Torr corresponds to the experimental parameters described in the beginning part of this paper and corresponds to that in figure-4. Here, the polarization rotation signal amplitude varies near linearly up to applied magnetic field of >±200 nT. In figure 5(b), the data for $^{87}$Rb atoms are shown in an expanded scale. It may be noted that the zero signal of the polarization rotation signal corresponds to finite magnetic field. The associated systemic error is found to be dependent on the various experimental parameters particularly degree of imbalance, position of locking, and depth of modulation applied to the RF. The noise amplitude in the polarization rotation signal is found to be ~0.5 nT and is dominated by the fluctuation in the locking of RF. It may be noted that the noise level is comparable to the CPT based magnetometer presented in ref-2. From the power density calculation, Schwindt *et al.* has estimated the sensitivity of the CPT based magnetometer to be 50 pT Hz$^{-1/2}$ at 10 Hz bandwidth with a shot-noise limited sensitivity of 1 pT Hz$^{-1/2}$ [2]. The demonstrated magnetometer is expected to have similar sensitivity due to use analogous technique. It is interesting to note that the VCO can be locked at a slight higher frequency then the common point described in fig-3. Under such a condition, the device cannot be used as atomic frequency standard as the RF frequency will be changed with the change in the magnetic field. However, the change in the RF will further increase the amplitude of the polarization rotation signal for same change in magnetic field as compared to VCO locked to the magnetic field insensitive position in fig-3. This can be realized by carefully looking in to fig-3. Therefore, the sensitivity of the magnetometer can surpass the limit of conventional CPT based magnetometer by utilizing the inherent option of synchronized magneto-assisted RF scanning and taking advantage of high SNR of the polarization rotation signal due to CPT states. The other conventional method of improving the sensitivity will be improvement in the SNR, working at narrower resonance width and improving the lock stability to the synthesized signal.

We have demonstrated a method of operation of a dual-purpose device for atomic frequency standard and magnetic field sensor. The difficulty of signal distortion of the transmitted signal due to CPT states being circumvented by using a synthesized signal generated from the transmission and polarization rotation signal by the CPT states. The most important aspect of the demonstrated magnetometer is elimination of any RF scanning in measurement of magnetic field. The magnetometer can detect the amplitude and orientation of magnetic field along the laser propagation direction and offer possibility of further improvement in sensitivity by inherent option of synchronized magneto-assisted RF scanning. The operations of the device in buffer gas and without buffer gas environment are compared.

**Acknowledgement:**

The authors are sincerely thankful to Dr. L.M. Gantayet, Director, BTDG for fruitful discussion and support during this work.

Note: Part of the work is filed for obtaining a patent [9]. The scientific outcomes of the applied patent and additional results are presented here.

**Figure Captions:**

**Figure-1:**
A VCSEL laser is frequency modulated by a voltage-controlled oscillator (VCO) to generate a bi-chromatic field. A controlled imbalance in the orthogonal circular polarization component is introduced by a polarizing beam splitter cube (PBS) and a quarter wave plate (WP). The tailored laser beam is made to interact with the atomic sample kept in a magnetic field controlled environment followed by monitoring the transmission along with the polarization rotation with the help of a PBS, pair photo diode (PD), signal processing units as a function of single photon as well as two photon detuning. The diode laser frequency is locked to the single photon resonance with a servo loop. The VCO frequency is locked to the synthesized signal generated from the transmission and polarization rotation due to CPT states.

**Figure-2:**
The transmission due to CPT states for -168 nT (red), ~0 T (black) and ~ +167 nT (blue) magnetic fields as a function of modulation frequency. The inset shows the expanded portion of the signal profile marked by the oval. The experiment corresponds to $^{87}$Rb atoms in presence of $N_2$ buffer gas at 25Torr. The angle of the quarter wave plate is ~$8^0$.

**Figure-3:**
The synthesized signal generated from the transmission and polarization rotation due to CPT state along with the polarization rotation signal for -168 nT (red), ~0 T (black) and ~ +167 nT (blue) as a function of modulation frequency. The polarization rotation signal has higher SNR due to signal extraction from almost zero background. Experimental parameters are same as described in figure-2.

**Figure-4:**
The VCO frequency locked to the common point of the synthesized signal illustrated in figure-3. The polarization rotation signal due to CPT states for various magnetic field values is found to be sensitive to the amplitude as well as orientation of the magnetic field along the laser propagation direction.

**Figure-5:**
The amplitude of the polarization rotation signal as a function of applied magnetic field. The signal amplitude is sensitive to both amplitude of the magnetic field as well as the orientation of the magnetic field along the laser propagation direction. **(a)** The empty square corresponds to the $^{85}$Rb D2 transition without any buffer gas with VCO is locked to the transmission due to CPT states. The angle of the quarter wave plate is ~$1^0$. The solid square represent the experimental data for $^{87}$Rb D1 transition in $N_2$ buffer gas at 25 Torr with VCO is locked to the combined signal as discussed in figure-4. Other parameters are same as in figure-2. **(b)** The expanded part of the buffer gas data.

**Figure-1:**

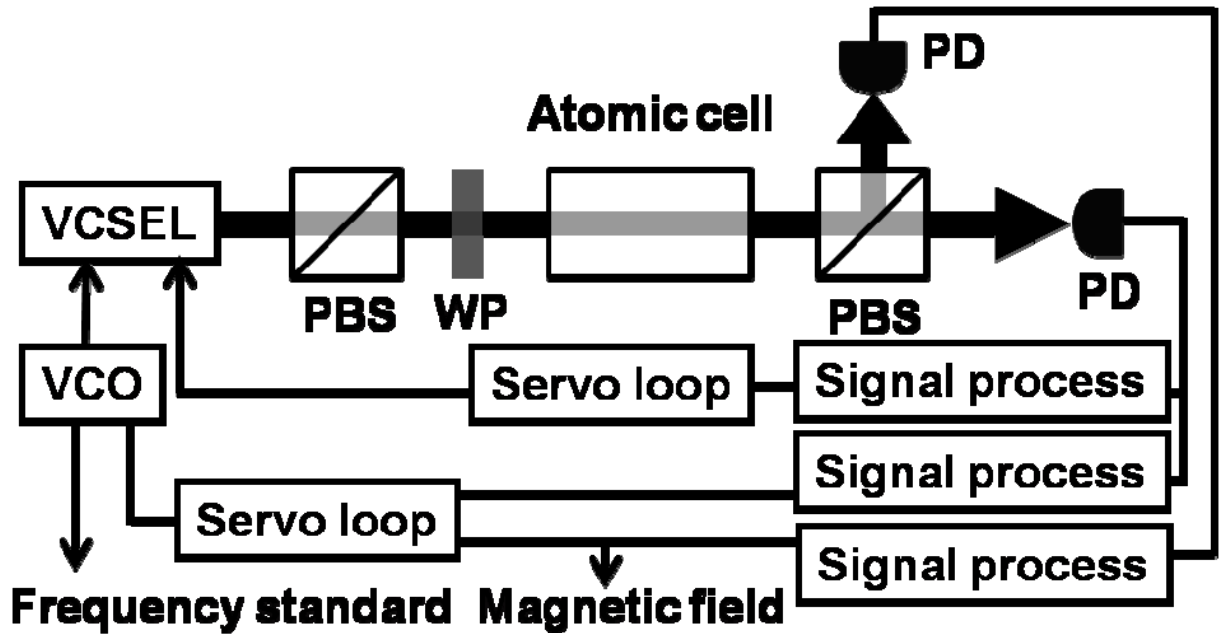

**Figure-2:**

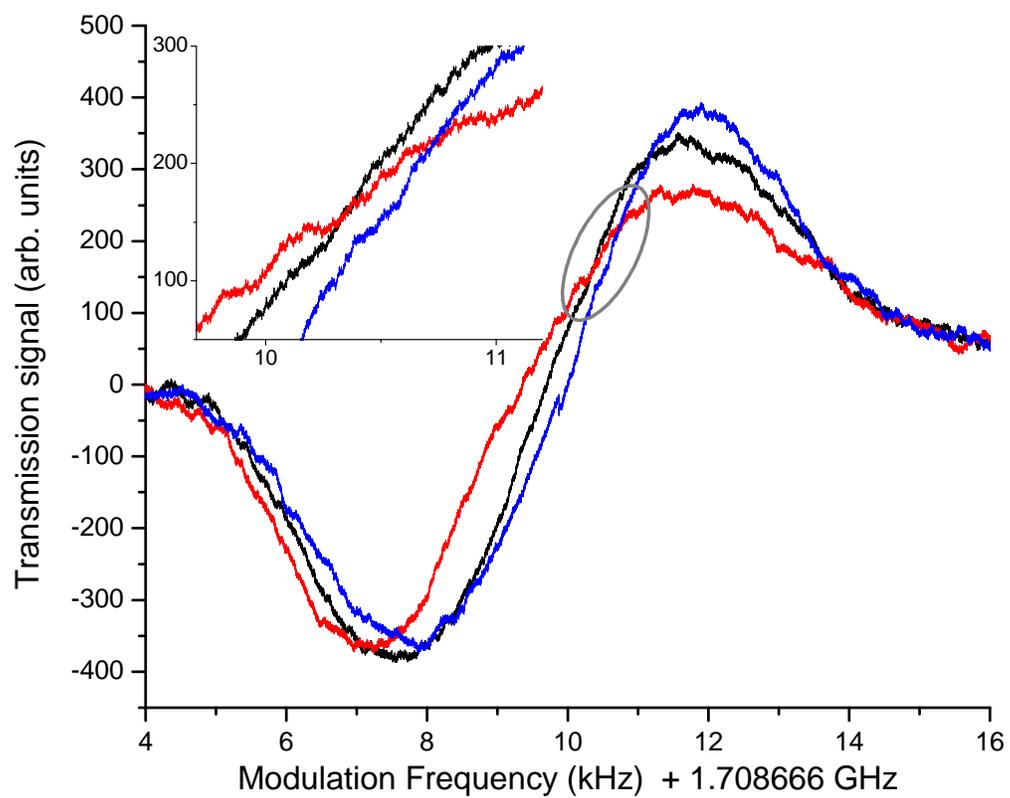

**Figure-3**

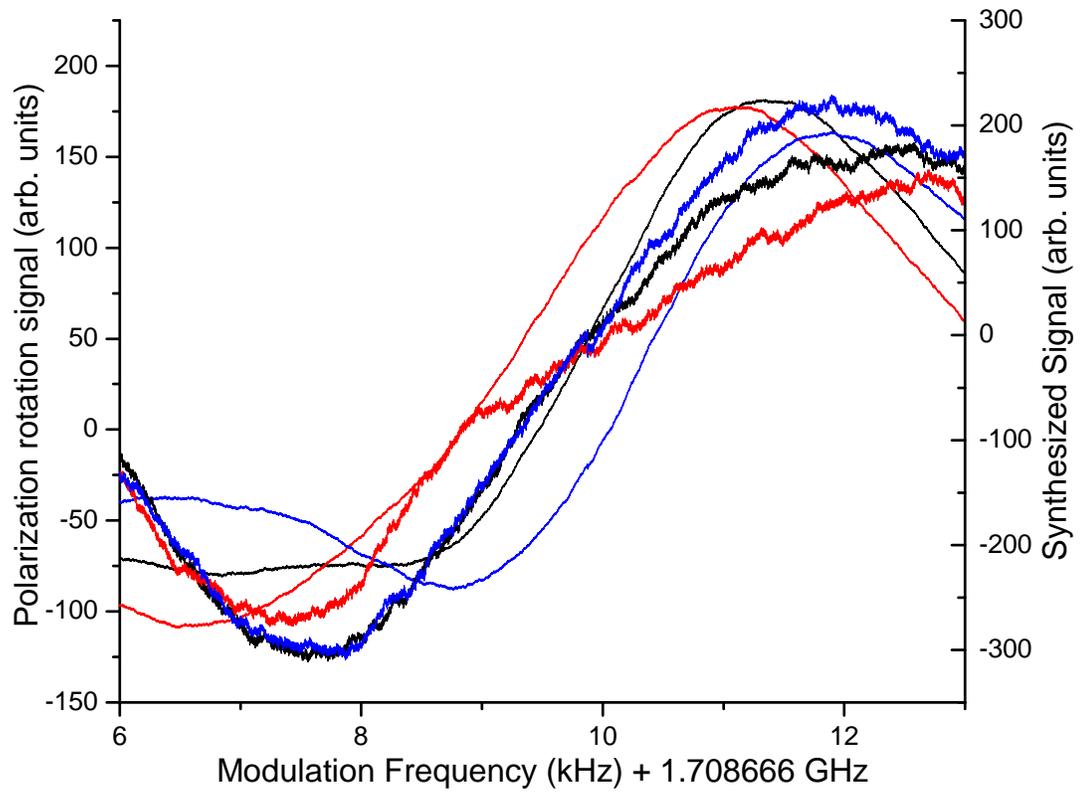

**Figure-4**

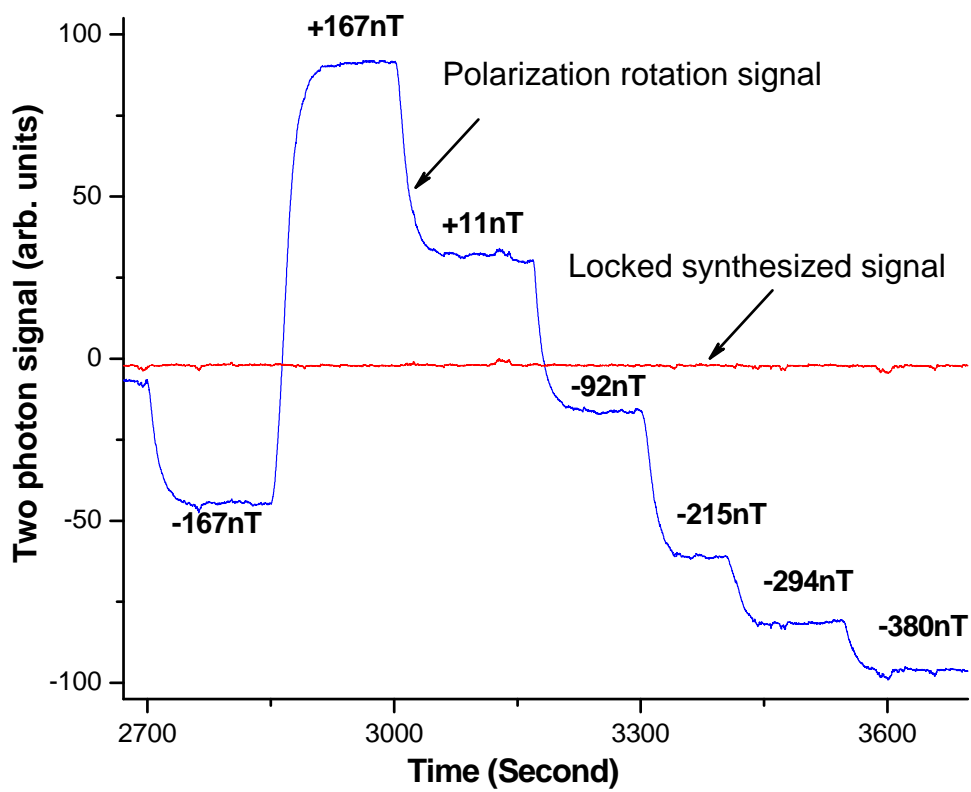

**Figure-5 (a)**

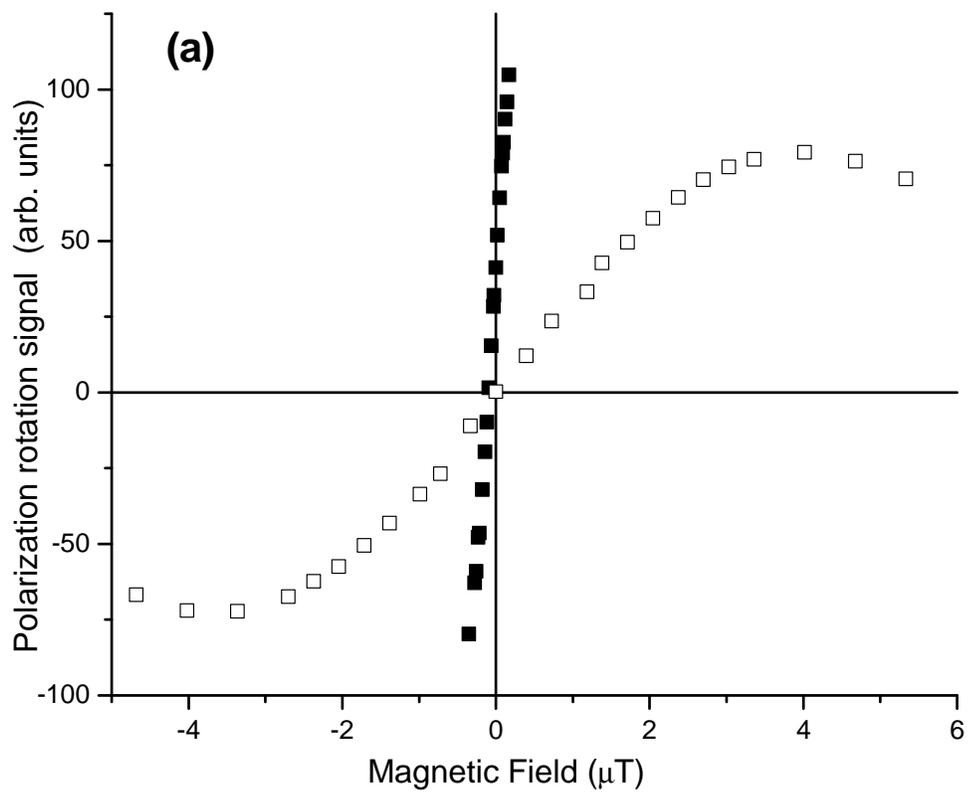

**Figure-5 (b)**

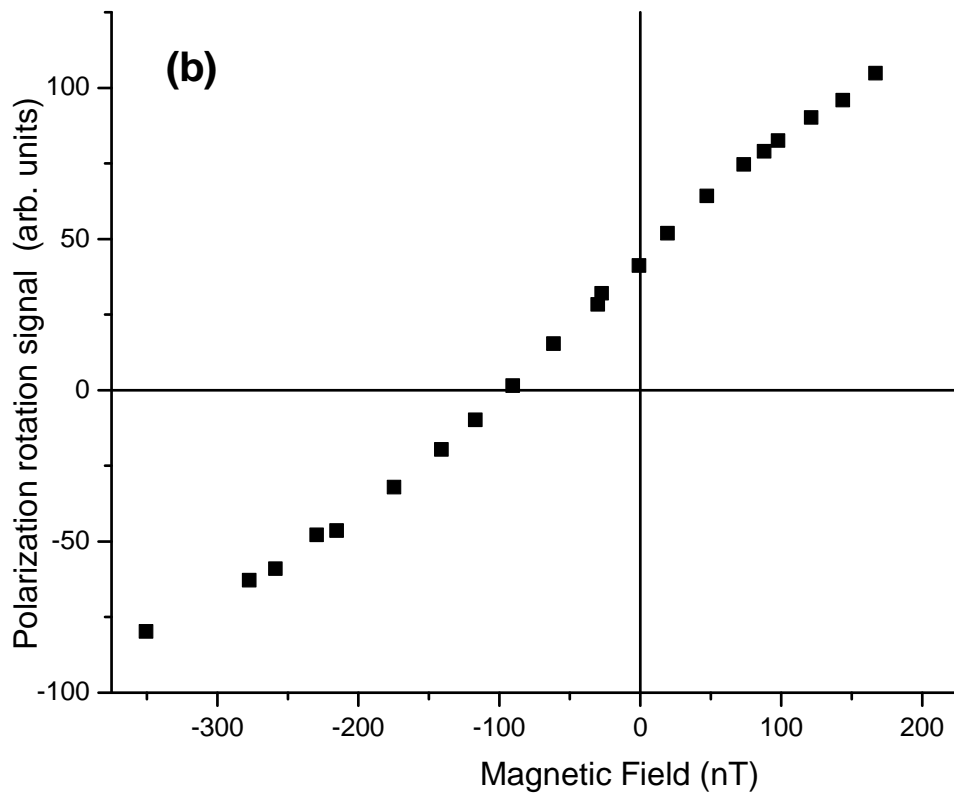